\documentclass[12pt]{article}
\usepackage{amsfonts}
\usepackage{amssymb}
\usepackage{amsthm}
\usepackage{graphicx}
\theoremstyle{plain}
\newtheorem{theorem}{Theorem}
\newtheorem{proposition}[theorem]{Proposition}
\newtheorem{lemma}[theorem]{Lemma}
\newtheorem{corollary}[theorem]{Corollary}

\theoremstyle{definition}
\newtheorem{definition}[theorem]{Definition}

\theoremstyle{remark}
\newtheorem{example}[theorem]{Example}
\newtheorem{examples}[theorem]{Examples}

\def\Asp{{\bf Asp}}

\newcommand{\maxtensor}{\otimes_{max}}
\newcommand{\mintensor}{\otimes_{min}}

\renewcommand{\hat}{\widehat}

\newcommand{\Hom}{\mbox{Hom}}
\newcommand{\Mor}{\mbox{Mor}}
\newcommand{\Pos}{\mbox{Pos}}
\newcommand{\FDNOrdLin}{\mbox{{\bf FDNOrdLin}}}
\newcommand{\FDOrdLin}{\mbox{{\bf FDOrdLin}}}

\renewcommand{\H}{{\cal H}}

\renewcommand{\phi}{\varphi}

\newcommand{\text}[1]{{\mbox{\rm #1}}}
\newcommand{\tr}{\text{Tr}}
\newcommand{\id}{\text{id}}

\newtheorem{Lemma}{Lemma}

\newcommand{\beq}{\begin{equation}}
\newcommand{\eeq}{\end{equation}}
\newcommand{\beqa}{\begin{eqnarray}}
\newcommand{\eeqa}{\end{eqnarray}}

\def\cc{{\cal C}}
\def\C{{\mathbf C}}
\def\R{{\mathbb R}}
\def\CC{{\mathbb C}}
\def\HH{{\mathbb HH}}

\def\iso{{\simeq}}

\newcommand{\ve}[1]{{\mathbf #1 }}

\def\id{{\rm id}}

 \title{Information processing in convex
  operational theories} \author{Howard Barnum\thanks{CCS-3:
    Information Sciences, Los Alamos National Laboratory, Los Alamos,
    NM, USA; Current affiliation: Perimeter Institute for Theoretical
    Physics, 31 Caroline St. N, Waterloo, Ontario, Canada {\tt
      hnbarnum@aol.com, hbarnum@perimeterinstitute.ca}} and Alexander
  Wilce\thanks{Department of Mathematical Sciences, Susquehanna
    University, Selinsgrove, PA, USA {\tt wilce@susqu.edu}}}
\date{June 29, 2008}
\begin{document}
\maketitle
\begin{abstract}
In order to understand the source and extent of the
greater-than-classical information processing power of quantum
systems, one wants to characterize both classical and quantum
mechanics as points in a broader space of possible theories. One
approach to doing this, pioneered by Abramsky and Coecke, is to
abstract the essential categorical features of classical and quantum
mechanics that support various information-theoretic constraints and
possibilities, e.g., the impossibility of cloning in the latter, and
the possibility of teleportation in both. Another approach, pursued
by the authors and various collaborators, is to begin with a very
conservative, and in a sense very concrete, generalization of
classical probability theory---which is still sufficient to
encompass quantum theory---and to ask which ``quantum"
informational phenomena can be reproduced in this much looser
setting. In this paper, we review the progress to date in this
second programme, and offer some suggestions as to how to link it
with the categorical semantics for quantum processes developed by
Abramsky and Coecke.
\end{abstract}
\section{Introduction}\label{intro}


The advent of quantum information theory has been accompanied by a
resurgence of interest in the convex (or ordered linear spaces)
framework for operational theories, as researchers seek to
understand the nature of information processing in increasingly
abstract terms, both in order to illuminate the sources of the
difference between the information processing power of quantum
theory and that of classical theory, and because quantum information has
occasioned renewed interest in foundational aspects of quantum
theory, often with the new twist that axioms or principles
concerning information processing are considered. A representative
(but by no means exhaustive) sample of work in this vein might
include the work of Hardy \cite{Hardy2001a,Hardy2001b}, D'Ariano
\cite{d'Ariano2006a}, and Barrett\cite{Barrett:2005a}.

At the same time, a fascinating and illuminating categorial approach
to the formulation of quantum physics has crystallized around the
notions of compact closed and dagger compact closed categories that
exhibit key features of quantum theory, but allow many other models as
well.  The main work along these lines has been done by Abramsky and
Coecke \cite{Abramsky:2004a}, by Selinger
\cite{Selinger2004a,Selinger2007a}, and by Baez \cite{Baez04a}.

In \cite{Abramsky:2004a}, Abramsky and Coecke established that many
of the most striking phenomena associated with quantum information
processing---notably, various forms of teleportation---arise much
more generally in any compact closed category, including, for
instance, the category of sets and relations. An important
observation here is that {\em the unit and co-unit defining a dual
object in such a category can be interpreted as a teleportation
protocol.} On the other hand, working in the much more concrete but
structurally much looser convex framework (in which essentially
arbitrary compact convex sets serve as abstract state spaces), our
coauthors (Jonathan Barrett and Matthew Leifer) and we have shown
([3]-[5]) that many of the same phenomena---in particular, many
aspects of entanglement, as well as no-cloning and no-broadcasting
theorems---are quite generic features of probabilistic models. In
this framework, the existence of a teleportation protocol is a
nontrivial constraint, moving one somewhat closer to quantum theory;
but even so, one can construct many models of teleportation---and
even of deterministic teleportation---that are neither classical
nor quantum. An important observation here is that a teleportation
protocol is {\em just a special case of conditioning}.

This paper reviews work by ourselves and various collaborators,
especially Jon Barrett, Matt Leifer, Oscar
Dahlsten, Leifer, and Ben Toner, on information processing in the ordered
linear spaces framework, and then proceeds to discuss how this
work may be related to the broad project of describing information-processing using
categories of processes.  The work reviewed shows that certain information-processing
properties which had sometimes been taken to be ``peculiarly
quantum,'' are actually common to all nonclassical theories in the
framework.  These include the existence of information about states
which cannot be obtained without disturbing them, and generalizations
of the quantum no-cloning and no-broadcasting theorems.

The impossibility of bit commitment has been suggested (for example
by Brassard \cite{Brassard:2005a} and by Fuchs \cite{Fuchs:2003}) as
a potential fundamental information-processing principle, shared by
classical and quantum mechanics, that might, in combination with
other principles, characterize quantum mechanics.  The other
principles proposed by Brassard and by Fuchs are the possibility of
secure secret key distribution, which is intimately connected with
no-cloning, no-broadcasting, and information-disturbance tradeoffs
and which, as we shall see, rules out classical theory, and the
impossibility of instantaneous signaling between systems (which is
built into the notion of composite system used in our version of the
ordered linear spaces framework).  We will also present some results
on bit commitment in our framework, to the effect that {\em all}
nonclassical theories that lack entanglement permit exponentially
secure bit commitment, and some results on how the presence in a
theory of certain kinds of entangled states can defeat the bit
commitment protocol we used for the unentangled case.  Closely
related states can permit teleportation, another
information-processing task whose possibility helps distinguish
between classes of nonclassical theories in our framework.  We
summarize some of our recent work with Barrett and Leifer on
multipartite composite systems and teleportation in the ordered
linear spaces framework.  In particular, we report necessary and
sufficient conditions for a composite of three systems to support a
conclusive teleportation protocol, and interesting sufficient
conditions for deterministic teleportation.

We then make some first steps towards a
category-theoretic formulation of our results. The abstract state
spaces that we consider naturally form a category; however, this is
far from being compact closed. For one thing, the dual of an
abstract state space is usually not, in any natural way, another
state space, but a different sort of beast altogether.
Nor are our categories generally monoidal: more typically, they
support a profusion of possible mechanisms for coupling systems,
bounded by a maximal (and maximally entangled) tensor product
$\maxtensor$, and a minimal (unentangled) product $\mintensor$. On
the other hand, there are various constructions by which one can
embed our category of state spaces in a larger category of {\em
processes} having a better behaved---in particular, monoidal and
self-dual---structure. Moving in the opposite direction, one can
focus on restricted categories that are, in a sense (made precise
below) ``closed under teleportation": as it happens, the entangled
state and effect corresponding to a correction-free teleportation
protocol are precisely the unit and co-unit of a duality.

Rather than building categories of processes from (categories of)
abstract state spaces, one might start from the opposite direction,
by treating categories of processes axiomatically. The idea that
processes should be given a central role in generalized probability
theory is certainly not new---indeed, several formulations of the
convex operational approach, notably those of Barrett
\cite{Barrett:2005a}, of D'Ariano \cite{d'Ariano2006a}, and the
operation algebras described in \cite{Barnum2003a}, take processes
(or ``operations") as fundamental. However, a category-theoretic approach
has
considerable and obvious advantages for framing any theory in which
processes are to be regarded as truly fundamental.

An important ingredient in operational theories is the idea that one
can {\em randomize} the preparation of a state, the choice of a
measurement, or, indeed, the selection of any sort of process. This
is reflected, for instance, in the convexity of state spaces and
spaces of effects. One way to capture this idea in a
category-theoretic framework is to consider categories {\em
enriched} over ordered linear spaces, or over abstract state spaces.
This paper ends with a sketch of this idea. We envision such
categorial formulations as a first step toward comparing the
necessary and/or sufficient conditions for various information
processing protocols or informational properties of theories,
obtained in the convex framework, with properties such as compact
closure, dagger compact closure, non-cartesianity and so forth that
have been used in the literature on categorial descriptions of
information processing. This section of the paper has benefited from
discussions with Abramsky, Armstrong, Coecke, and others and may be
viewed as describing work early-in-progress in collaboration with at
least some of them.

\section{Abstract State Spaces}

By an {\em abstract state space}, we mean a pair $(A,u_{A})$ where
$A$ is a finite-dimensional ordered real vector space, with positive
cone $A_{+}$, and where $u_{A} : A \rightarrow {\mathbb R}$ is a
distinguished linear functional, called the {\em order unit}, that
is {\em strictly} positive on $A_{+} \setminus \{0\}$. A state is
{\em normalized} iff $u_{A}(\alpha) = 1$. We write $\Omega_{A}$ for
the convex set of normalized states in $A_{+}$. By way of
illustration, if $A$ is the space ${\mathbb R}^{X}$ of real-valued
functions on a set $X$, ordered pointwise on $X$, with $u_{A}(f) =
\sum_{x \in X} f(x)$, then $\Omega_{A} = \Delta(X)$, the simplex of
probability weights on $X$. If $A$ is the space ${\cal L}(\H)$ of
hermitian operators on a (finite-dimensional) complex Hilbert space
$\H$, with the usual operator ordering (whose positive cone is the positive
semidefinite operators), and if $u_{A}(a) = \tr(a)$,
then $\Omega_{A}$ is the set of density operators on $\H$. On any
abstract state space $A$, there is a canonical norm (the {\em base
norm}) such that for $\alpha \in A_{+}$, $\|\alpha\| =
u_{A}(\alpha)$. For ${\mathbb R}^{X}$, this is just the norm on $X$;
for ${\cal L}(\H)$, it is the trace norm.

Events (e.g., measurement outcomes) associated with an abstract
state space $A$ are represented by {\em effects}, i.e.,  positive
linear functionals $a \in A^{\ast}$, with $0 \leq a \leq u_{A}$ in
the dual ordering. Note that $0$ and $u_{A}$ are, by definition, the
least and greatest effects. If $\alpha$ is a {\em normalized} state
in $A$---that is, if $u_{A}(\alpha) = 1$---then we interpret
$a(\alpha)$ as the {\em probability} that the event represented by
the effect $a$ will occur if measured. Accordingly, a discrete {\em
observable} on $A$ is a list $(a_1,...,a_n)$ of effects with $a_1 +
a_2 + \cdots + a_n = u_{A}$. We represent a physical process with
initial state space $A$ and final state space $B$ by a positive
mapping $\tau : A \rightarrow B$ such that, for all $\alpha \in
A_{+}$, $u_{B}(\tau(\alpha)) \leq u_{A}(\alpha)$---equivalently,
$\tau$ is norm-contractive. We can regard $\|\tau(\alpha)\| =
u_{B}(\tau(\alpha))$ as the probability that the process represented
by $\tau$ takes place in initial state $\alpha$; this {\em event} is
represented by the effect $u_{B} \circ \tau$ on $A$.

It is important to note that, in the framework just outlined, the
state space $A$ and its dual space $A^{\ast}$ have (in general)
quite different structures: $A$ is a {\em cone-base space} (a.k.a. 
{\em base-norm space}), i.e., an
ordered space with a preferred base, $\Omega_A$, for $A_+$, while
$A^{\ast}$ is an {\em order-unit} space, i.e., an ordered space with
a preferred {\em element} in its positive cone. Indeed, the spaces
$A$ and $A^{\ast}$ are generally not even isomorphic as ordered
spaces. Where there exists an order-isomorphism (that is, a
positive linear mapping with positive inverse) between $A$ and
$A^{\ast}$, we shall say that $A$ is {\em weakly self-dual}. Where
this isomorphism induces an inner product on $A$ such that $A_{+} =
\{ b \in A | \langle b, a \rangle \geq 0 \ \forall a \in A_{+}\}$,
we say that $A$ is {\em self-dual}. Finite dimensional quantum and
classical state spaces are self-dual in this sense. A celebrated
theorem of Koecher and of Vinberg [8,9] tells us that if $A$ is an
irreducible, finite-dimensional self-dual state space, and if the
group of affine automorphisms of $A_{+}$ acts transitively on the
interior of $A_{+}$, then the space $\Omega_{A}$ of normalized
states is affinely isomorphic to the set of density operators on an
$n$-dimensional Hilbert space over $\R, \CC$, or $\HH$, or to a ball, or to the set of $3
\times 3$ trace-one positive matrices over the octonions.

\section{Composite Systems}

For our purposes, it will be convenient to identify the tensor
product, $A \otimes B$, of two state spaces with the space
$B(A^{\ast}, B^{\ast})$ of bilinear forms on $A^{\ast} \times
B^{\ast}$, interpreting the pure tensor $\alpha \otimes \beta$ of
states $\alpha \in A, \beta \in B$ as the form given by $(\alpha
\otimes \beta)(f,g) = f(\alpha)g(\beta)$ where $f \in A^{\ast}, g
\in B^{\ast}$. We call a form $\omega \in A \otimes B$ {\em
positive} iff $\omega(a,b) \geq 0$ for all $(a,b) \in A^{\ast}_{+}
\times B^{\ast}_{+}$. If $\omega$ is positive and $\omega(u_A, u_B)
= 1$, then $\omega(a,b)$ can be interpreted as a joint probability
for effects $a \in A^{\ast}$ and $b \in B^{\ast}$. Conversely, one
can show (see [4] and [10]) that any assignment of joint
probabilities consistent with a no-signalling requirement must be
bilinear. Thus, the most general model of a composite of $A$ and $B$
consistent with such a requirement, is the space $A \otimes B$,
ordered by the cone of all positive forms, and with order unit given
by $u_{A} \otimes u_{B} : \omega \mapsto \omega(u_{A}, u_{B})$. This
gives us an abstract state space, which we term the {\em maximal
tensor product} of $A$ and $B$, and denote $A \maxtensor B$. At the
other extreme, we might wish to allow only product states $\alpha
\otimes \beta$, and mixtures of these, to count as bipartite
(normalized) states. This gives us the {\em minimal tensor product},
$A \mintensor B$. These coincide if $A$ and $B$ are classical --
that is, if $\Omega_{A}$ and $\Omega_{B}$ are simplices [11]; in
general, however, the maximal tensor product allows many more states
than the minimal. A state in $\Omega_{A \maxtensor B}$ not belonging
to $\Omega_{A \mintensor B}$ is {\em entangled}.

More generally, we define a {\em composite} of $A$ and $B$ to be
{\em any} state space $AB$ consisting of bilinear forms on $A^{\ast}
\times B^{\ast}$, ordered by a cone $AB_{+}$ of positive forms
containing every product state $\alpha \otimes \beta$, where $\alpha
\in \Omega_{A}$ and $\beta \in \Omega_{B}$---equivalently, $AB$ is
a composite iff $A \mintensor B \leq AB \leq A \maxtensor B$ (where,
for abstract state spaces $A$ and $B$, $A \leq B$ means that $A$ is
a subspace of $B$, that $A_{+} \subseteq B_{+}$, and that $u_{A}$ is
the restriction of $u_{B}$ to $A$.) More generally still, a
composite of $n$ state spaces $A_1,...,A_n$ is a state space $A$ of
$n$-linear forms on $A_1^{\ast} \times \cdots \times A_{n}^{\ast}$,
ordered by any cone of positive forms containing all product states.


\section{Information-disturbance tradeoffs}
\label{subsec: information-disturbance}

With Barrett and Leifer, we have shown (as described in
\cite{Barrett:2005a}) that in nonclassical theories, the only
information that can be obtained about the state without disturbing
it is inherently classical information---information about which of
a set of irreducible direct summands of the state cone the state
lies in. Call a positive map $T : A \rightarrow A$ {\em
nondisturbing} on state $\omega$ if $T(\omega) = c_{\omega} \omega$
for some positive constant $c_{\omega}$ that in principle could
depend on the state. Say such a map is {\em nondisturbing} if it is
nondisturbing on all pure states.\footnote{ Of course if we
condition on information obtained, this definition permits mixed
states to be disturbed by a nondisturbing map---that can be viewed
as something like an inevitable ``epistemic'' disturbance associated
with obtaining information.}
A norm-nonincreasing map nondisturbing in
this sense is precisely the type of map that can appear
associated with some measurement outcome in an operation that,
averaged over measurement outcomes, leaves the state (pure or not) unchanged.

A cone $C$ in a vector space $V$
is a {\em direct sum} of cones $D$ and $E$ if $D$ and $E$ span
disjoint (except for $0$) subspaces of $V$, and every element of
$C$ is a positive combination of vectors in $D$ and $E$.  A cone is
irreducible if it is not a nontrivial direct sum of cones.  Every
finite-dimensional cone is uniquely expressible as a
direct sum $C = \oplus_i C_i$ of irreducible cones $C_i$.  Information
about which of the summands a state is in should be thought of as ``inherently
classical'' information about the state.

\begin{theorem}
The nondisturbing maps on a cone that is a sum $C =
\oplus_i C_i$ of irreducible $C_i$, are precisely the maps
$M = \sum_i c_i \id_i$, where $\id_i$ is the identity operator on the
summand $V_i$ and the zero operator elsewhere, and $c_i$ are arbitrary
nonnegative constants.
\end{theorem}

So for a nondisturbing map, $c_\omega$ can depend only on
the irreducible component a state is in.  That is, the
fact that a nondisturbing map has occured can give us no
information about the state within an irreducible component: in other
words, as claimed, only inherently classical information is contained
in the fact that a nondisturbing map has occured.

The existence of information that cannot be obtained without
disturbance is often taken to be the principle underlying the
possibility of quantum key distribution, so the fact that it is
generic in nonclassical theories in the framework leads us (with
Barrett and Leifer) to conjecture that secure key distribution,
given an authenticated public channel, is possible in all
nonclassical models.

\subsection{No-cloning and no-broadcasting theorems}
\label{subsec: no-cloning and no-broadcasting}
The security of quantum key distribution is also often ascribed to the
quantum no-cloning or no-broadcasting theorem---certainly no-cloning is
at least {\em necessary} for security.  A map $T: A \rightarrow
A \otimes A$
{\em clones} a state $\omega$ if $T(\omega) = \omega \otimes \omega$.  A {\em set}
$S$ of normalized states can be (deterministically) cloned if there is
a single dynamically allowed map $T$ that clones every $\omega \in S$.

No-cloning can be closely related to the information-disturbance
principle, by an argument introduced in the quantum context but that
generalizes to our setting, since if two non-identical states in the
same irreducible component of a cone could be cloned, we could---by,
for instance, doing an informationally complete measurement on the
clone---obtain information about which state we have without
disturbing it, contradicting our information-disturbance theorem.

In quantum mechanics, only orthogonal sets of states---sets $S$ such
that for all pairs $\rho, \sigma \in S$, $\rho \sigma = 0$---can be
cloned \cite{Barnum:1996}.  As a special case of this, in a classical
probability theory with a finite sample space, sets 
containing properly mixed states
(distributions) cannot be cloned (except for singletons).  Because of this, and because it is
natural to consider {\em commuting} rather than mutually orthogonal
sets of density matrices to be ``classical subsets'' of the quantum
states of a system, \cite{Barnum:1996} introduced the notion of {\em
  broadcasting} in order to better pick out classical subsets of the
state spaces of quantum systems.  A map $T: A \rightarrow A \otimes A$
{\em broadcasts} a state $\omega$ if both marginals of $T(\omega)$ are
equal to $\omega$; thus this notion allows correlation, or even
entanglement, in the broadcast state.  This is to be contrasted with
the mixed-state extension of the notion of cloning for which we used
the term ``cloning'' above, which produces a product state.  Of
course, the notion of broadcasting also extends, in a different way,
the notion of cloning pure states, since it reduces to cloning on pure
states.  A set $S$ of states is {\em broadcastable} if there is a
norm-preserving dynamical map $T$ that broadcasts all the states in
$S$ (i.e. the {\em same} map broadcasts all the states).

The {\em no-broadcasting theorem} \cite{Barnum:1996} asserts that it
is precisely the mutually commuting sets of quantum states that can
be broadcast using completely positive maps.  Recently, with Barrett
and Leifer we have shown \cite{Barnum:2006} the following:

\begin{theorem}
In an arbitrary convex operational theory
in our framework, a set $S \subseteq \Omega$ of states is
broadcastable if, and only if, it is contained in a simplex $\Delta
\subseteq \Omega$ whose vertices are distinguishable by a single
measurement.  For each positive  map $B: V \rightarrow V \otimes V$, the set of states
it broadcasts is {\em precisely} such a simplex.
\end{theorem}

This combines Theorems 2 and 3 of \cite{Barnum:2006}.  It can be
interpreted as saying that broadcastable sets of states are classical
sets of states---but the sort of classicality involved is different from
the {\em inherent} classicality of the information that can be
obtained without disturbance.

The proof of the theorem uses a generalized no-cloning theorem, also
proved in \cite{Barnum:2006}, to the effect that a set of states is
clonable if, and only if, the states in it are all distinguishable
from each other simultaneously via a one-shot measurement.  Given
this, proving Theorem 2 reduced to proving that a broadcastable set
of states is contained in (and the states broadcast by $B$ are {\em
precisely}) the convex hull (necessarily a simplex) of a clonable
set of states.  The proof of the no-cloning result is essentially to
show that if one can clone a set of states, one can distinguish them
by repeatedly cloning to create many independent copies, performing
an informationally complete measurement on each copy, and using the
statistics of the measurement results to identify the state.
Conversely if one can distinguish the states, one can clone them
using a map that, conditional on distinguishing state $\omega$,
prepares $\omega \otimes \omega$.  More precisely: for any $\omega$
there is a norm-nonincreasing positive map $Prep_{\omega}$ that
prepares $\omega$, i.e.  outputs $\omega$ no matter what normalized
state goes in.  The cloning map is $\sum_i Prep_{\omega_i \otimes
\omega_i} \circ T_i$, where $\{T_i\}$ are a set of maps such that
the effects $u \circ T_i$ are a measurement distinguishing the
$\rho_i \in S$; such a measurement must exist by our assumption the
$\rho_i$ were one-shot distinguishable, and we assumed as part of
our general framework that every effect has at least one associated
map $T_i$.  One immediately sees that this map that clones the
$\omega_i$ will also broadcast any state in the convex hull of the
$\omega_i$, giving us the easy direction of the generalized
no-broadcasting theorem.

\section{Nonuniqueness of extremal decomposition,
and bit commitment in unentangled theories}
\label{subsec: unentangled bit commitment}

Quantum theory has mixed states whose representation as a convex
combination of pure states is not unique.  So do {\em all}
nonclassical theories: uniqueness of the decomposition of mixed states
into pure states is an easy characterization--sometimes used as a
definition---of simplices (see, for example, the proof in
\cite{Barrett:2005}).  While we are not aware of any quantum information
processing task whose possibility is directly traced to the non-unique
decomposability of mixed states into pure, this was certainly proposed
as a possible basis for quantum bit commitment schemes, though (as
shown in \cite{Bennett84a} for their proposed scheme, and in
\cite{Mayers97a,Lo97a} for more elaborate schemes) these schemes do
not work because of entanglement.

In \cite{BDLT2008a} it is shown that the existence of bit commitment
protocols is universal in nonclassical theories in the convex sets
framework, provided that the
tensor products used do not permit entanglement. Consider a theory
generated by a finite set $\Sigma$ of ``elementary'' systems modeled
by finite-dimensional abstract state spaces, containing at least one
nonclassical system, and closed under the minimal, or separable,
tensor product, which we write with the ordinary tensor product
symbol $\otimes$.

\noindent
{\em The protocol}.
Let a system have a non-simplicial, convex, compact state space
$\Omega$ of dimension $d$, embedded as the base of a cone
of unnormalized states in a vector space $V$ of dimension $d+1$.
The  protocol uses a state $\mu$ that has two distinct decompositions into
finite disjoint sets $\{\mu^0_i\}, \{\mu^1_j\}$ of exposed states,  that is,
\beqa
\label{2Decomp}
\omega = \sum_{i = 1}^{N_0} p_i^0 \mu^0_i = \sum_{j=1}^{N_1} p_j^1 \mu^1_j,
\eeqa

A state $\mu_i^b$ is {\em exposed} if there is a measurement outcome
$a_i^b$ that has probability $1$ when, and only when, the state is
$\mu_i^b$.  We call this outcome the {\em distinguishing effect} for 
$\mu_i^b$.  The protocol exists for all nonclassical systems because,
as we show, any non-simplicial convex set of affine dimension $d$ always
has a state $\omega$ with two decompositions (as above), into {\em disjoint}
set of states whose total number $N_0 + N_1$ is $d+1$ (the disjointness
and the bound on cardinality are used in the proof of exponential security).

In the honest protocol, Alice first decides on a bit $b \in \{0,1\}$
to commit to.  She then draws $n$ samples from $p^b$, obtaining a
string $\ve{x} = (x_1,x_2,\ldots,x_n )$.  To commit, she sends the
state $\mu^b_{\ve{x}} = \mu^b_{x_1} \otimes \mu^b_{x_2} \otimes \ldots
\otimes \mu^b_{x_n}$ to Bob.  To reveal the bit, she sends $b$ and
$\ve{x}$ to Bob.  Bob measures each subsystem of the state he has.  On
the $k$-th subsystem, he performs a measurement, (which will depend on
$b$) containing the distinguishing effect for $\mu^b_{x_k}$ and
rejects if the result is not the distinguishing effect.  If he obtains
the appropriate distinguishing effect for every system, he accepts.
The protocol is perfectly sound (if Alice
is honest, Bob never accuses her of cheating and always obtains the
correct bit), perfectly hiding (if Alice is honest,
Bob cannot gain any information about the bit until Alice reveals it),
and has an exponentially low probability of Alice's successfully cheating.

\section{Conditioning and teleportation protocols}

If $AB$ is a composite of state spaces $A$ and $B$, we can define,
for any normalized state $\omega \in AB_{+}$ and any effect $a \in
A$, both a {\em marginal} state $\omega_{A}( - ) = \omega(-, u_{B})$
and a {\em conditional} state $\omega_{B|a}(b) =
\omega(a,b)/\omega_{A}(a)$ (with the usual proviso that if
$\omega_{A}(a) = 0$, the conditional state is also $0$). We shall
also refer to the partially evaluated state $\omega_{B}(a) :=
\omega( a, - )$ as an {\em un-normalized} conditional state.

More generally, if $A$ is a composite of state spaces $A_1,...,A_n$,
with order-units $u_1,...,u_n$, then for all subsets $J \subseteq
\{1,...,n\}$, and all $a := (a_i) \in \otimes_{i \not \in J}
A_{i}^{\ast}$, we can define an un-normalized conditional state --
that is, a partially evaluated
 state---$\omega_{J}^{a}$, a $|J|$-linear form on $\Pi_{j
\in J} A_{j}^{\ast}$. We define the {\em $J$-th subsystem} to be the
the ordered space spanned by the cone generated by these conditional
states, with order unit $u_{J} := \otimes_{j \in J} u_{j}$. We call
$A$ a {\em regular} composite iff it is closed under taking products
of such multi-partite conditional states. All state spaces
constructed from a single, associative, bilinear product are
regular, but one can also build regular composites using ``mixed"
constructions. For instance, it is not difficult to show that $A
\mintensor (B \maxtensor C)$ is a regular composite of $A, B$ and
$C$. An example of a non-regular composite is $(A \mintensor B)
\maxtensor (C \mintensor D)$ where $A, B, C$ and $D$ are four copies
of a weakly self-dual, but non-classical, state space.

A state $\omega \in AB$ gives rise to a positive operator
$\hat{\omega} : A^{\ast} \rightarrow B$, given by
$\hat{\omega}(a)(b) = \omega(a,b)$. We can regard $\hat{\omega}(a)$
as an ``un-normalized" conditional state. As a partial converse, any
positive operator $\psi : A^{\ast} \rightarrow B$ with $\psi(u_{A})
\in \Omega_{B}$---that is, with $\psi^{\ast}(u_{B}) := u_{B} \circ
\psi = u_{A}$---corresponds to a state in the maximal tensor
product $A \maxtensor B$. Dually, any effect $f \in (AB)^{\ast}$
yields an operator $\hat{f} : A \rightarrow B^{\ast}$, given by
$\hat{f}(\alpha)(\beta) = f(\alpha \otimes \beta)$; and any positive
operator $\phi : A \rightarrow B^{\ast}$ with $\phi(\alpha) \leq
u_{B}$ for all $\alpha \in \Omega_{A}$---that is, with $\|\phi\|
\leq 1$---corresponds to an effect in $(A \mintensor B)^{\ast}$. We
have the following result (easily verified by checking that it holds
for elementary tensors):

\begin{Lemma} Let $ABC$ be a regular composite. If $f$ is an effect in $(AB)^{\ast}$ and $\omega$ is a state
in $BC$, then, for any $\alpha \in A$,
\begin{equation}(\alpha\otimes \omega)^{B}_{f} = \|\hat{\omega} ( \hat{f}(\alpha)) \|\hat{\omega}
(\hat{f}(\alpha)).\end{equation} \end{Lemma}

If $ABC$ is in a state $\alpha \otimes \omega$, with $\alpha$
unknown, then conditional on securing measurement outcome $f$ on $A
\otimes B$, the state of $C$ is, up to normalization, a {\em known
function of} $\alpha$. We call this {\em remote evaluation.} This is
very like a teleportation protocol. Indeed, suppose that $C$ is a
copy of $A$, and that $\eta : A \rightarrow C$ is a specified
isomorphism allowing us to match up states in the former with those
in the latter:

\begin{definition} With notation as above, $(f,\omega)$ is a (one-outcome, post-selected) {\em teleportation
protocol} iff there exists a positive, norm-contractive {\em
correction map} $\tau : C \rightarrow C$ such that, for all $\alpha
\in A$, $\tau(\alpha \otimes \omega)^{C}_{f} =
\eta(\alpha)$.\footnote{One could also allow protocols in which the
correction has a nonzero probability to fail. For details, see
\cite{Barnum:2008}.}
\end{definition}

By Lemma 1, the un-normalized conditional state of $\alpha \otimes
\omega$ is exactly $\hat{\omega}(\hat{f}(\alpha))$. If we let $\mu
:= \hat{\omega} \circ \hat{f}$, the normalized conditional state can
be written as $\mu(\alpha)/u(\mu(\alpha))$. Thus, $(f,\omega)$ is a
teleportation protocol iff there exists a norm-contractive mapping
$\tau$ with $(\tau \circ \mu)(\alpha) = \|\mu(\alpha)\|\eta$ for all
$\alpha \in \Omega_{A}$.

\begin{theorem}[\cite{Barnum:2008}] With notation as above, $(f,\omega)$ is a teleportation protocol iff $\mu : = \hat{\omega} \circ \hat{f}$ is proportional
to an isomorphism $(A,u_{A}) \simeq (C,u_{C})$; in this case, the
correction $\tau : (C,u_{C}) \simeq (C,u_{C})$ is also an
isomorphism.
\end{theorem}

Henceforth, we simply identify $C$ with $A$, suppressing $\eta$.
Note that if $(f,\omega)$ is a teleportation protocol on a regular
composite $A B A$ of $A$, $B$ and (a copy of) $A$, then, as $f$
lives in $(AB)^{\ast} \leq (A \mintensor B)^{\ast}$ and $\omega$
lives in $BA \leq B \maxtensor C$, one can also regard $(f,\omega)$
as a teleportation protocol on $A \mintensor (B \maxtensor A)$.

\begin{theorem}[\cite{Barnum:2008}] $A \mintensor (B \maxtensor A)$ supports a conclusive teleportation protocol iff $A_1$ is order-isomorphic to the
range of a compression (a positive idempotent mapping) $P :
A_2^{\ast} \rightarrow A_1$. \end{theorem}

\begin{corollary} If $A$ can be teleported through a copy of
itself, then $A$ is weakly self-dual. \end{corollary}

In order to {\em deterministically} teleport an unknown state
$\alpha \in A$ through $B$, we need not just one entangled effect
$f$, but an entire observable's worth. Here, we specialize to the
case in which $ABC = A \mintensor (B \maxtensor C)$:

\begin{definition} A {\em deterministic teleportation protocol} for $A$ through $B$ consists of
an observable $E = (f_1,...,f_n)$ on $A \otimes B$ and a state
$\omega$ in $B \otimes A$, such that for all $i = 1,...,n$, the
operator $\hat{f_i} \circ \hat{\omega}$ is 
invertible via a dynamically allowed map.\end{definition}

The following result provides a sufficient condition (satisfied,
e.g., by any state space $A$ with $\Omega_{A}$ a regular polygon)
for such a protocol to exist.

\begin{theorem}[\cite{Barnum:2008}] Let $A = B$. Suppose that $G$ is a finite group acting transitively on the pure states of $A$,
and let $\omega$ be a state such that $\hat{\omega}$ is a
$G$-equivariant isomorphism. For all $g \in G$, let $f_{g} \in (A
\maxtensor A)^{\ast}$ correspond to the operator
\[\hat{f}_{g} = \frac{1}{|G|} \hat{\omega}^{-1} \circ g.\]
Then $E = \{f_{g} | g \in G\}$ is an observable, and $(E, \omega)$
is a deterministic teleportation protocol.\end{theorem}

\section{Categories of Abstract State Spaces}

If an abstract state space and its dual ``effect space" provide an
abstract probabilistic model, one should like to say that a
probabilistic {\em theory} is a class of such models, closed under
appropriate operations. To make this systematic, one should consider
categories of state spaces. Let $\Asp$ denote the category whose
objects are finite-dimensional abstract state spaces $(A, u_{A})$,
and whose morphisms are norm-contractive positive linear mappings.
This category has a preferred object $I = {\mathbb R}$, ordered as
usual, with $u_I = 1$ and $\Omega_I = \{1\}$. For any $A$ in $\Asp$,
there is a preferred morphism, namely $u_A$, from $A$ to $I$.
(Indeed, the mappings $\tau \mapsto u_{A} \circ \tau$ define a
natural transformation $\Asp(-,A) \rightarrow \Asp(-,I)$.) We can
model effect spaces, i.e., dual state spaces, by Hom sets
$\Asp(A,I)$. However, as remarked above, there is no natural {\em
internal} duality for $\Asp$; nor is $\Asp$ naturally a monoidal
category, owing to the the existence of {\em two} canonical tensor
products $\maxtensor$ and $\mintensor$. These interact in a way that
will be familiar to linear-logicians, namely, for any state spaces
$A$, $B$ and $C$, there is a canonical embedding
\[ A \mintensor (B \maxtensor C) \leq (A \mintensor B) \maxtensor
C.\]

Thus, we can regard $(\Asp, \mintensor, \maxtensor)$ as a {\em
linearly distributive} category [6] (albeit without negation). As to
duality, there are various constructions whereby a useful
self-duality can be supplied.   Applied to $\Asp$, these result in
what may be regarded as categories of {\em process spaces}. To take
the simplest case, consider the category ${\C}^{2} = {\C} \times
{\C}$, i.e., the category whose objects are ordered pairs $(A,B)$ of
state spaces in $\Asp$, and  in which
\[{\C}^2((A,B),(C,D)) =
{\C}(C,A) \times {\C}(B,D),\] with composition defined in the
obvious way.\footnote{This is essentially the category ${\C}^{d}$
described in \cite{Hyland-Schalk:2003}.  Another possibility 
for a category of processes would be
to apply the Int \cite{Hyland-Schalk:2003} (or ``GoI"
\cite{Abramsky:2004a}) construction to $\Asp$.  We shall not pursue
this here.} The idea is that the pair
$(A,B)$ represents a space of possible processes from $A$ to $B$,
and that a pair $f : C \rightarrow A$, $g : B \rightarrow D$ takes a
process $\tau : A \rightarrow B$ to the process $g \circ \tau \circ
f$. Indeed, the functor $(A,B) \mapsto {\C}(A,B)$ endows ${\C}^2$
with exactly this interpretation. Thus, $(A,I)$ encodes $A^{\ast}$,
while $(I,A)$ encodes $A$. Thus, ${\C}^2$ allows us to consider
state spaces and effect spaces on an equal footing. Moreover,
${\C}^{2}$ has a natural self-duality, given by $(A,B)^{\ast} =
(B,A)$ and, for $(f,g) \in {\C}^2((A,B), (C,D))$,
\[(f,g)^{\ast} = (g,f) \in {\C}^2((C,D)^{\ast}, (A,B)^{\ast}) =
{\C}^2((D,C),(B,A)).\] Where ${\C}$ is closed under both maximal and
minimal tensor products (in particular, for ${\C} = \Asp$), the
category ${\C}^{2}$ has a natural symmetric monoidal structure given
by
\[(A,B) \otimes (C,D) = (A \mintensor B, C \maxtensor D).\]
Note that we then have the expected identity \[ ((A,B)^{\ast}
\otimes (C,D)^{\ast})^{\ast} = (A \maxtensor B, C \mintensor D).\]

Rather than enlarging the category $\Asp$, one can also look within
it for subcategories with a desirable structure. Let ${\C}$ be a
subcategory of $\Asp$. If $A$ and $B$ are state spaces in ${\C}$, let
us agree that, as in Definition 2, a {\em teleportation protocol}
for $A$ {\em through} $B$ consists of (i) a regular composite $ABA
\in {\C}$; and (ii) a pair $f \in {\C}(AB,I)$, $\omega \in BA =
{\C}(I,BA)$ such that, for some {\em correction} $\tau \in {\C}(A,A)$,
\[\tau((f \otimes -) ( - \otimes \omega)) = \id_{A}.\]
We shall say that this protocol is {\em correction free} if $\tau$
can be taken to be the identity morphism $\id_{A}$.
This may look familiar. Recall that a {\em dual} for an object $A$
in a symmetric monoidal category $({\C},\otimes)$ is an object $B$,
together with morphisms $\eta_A : I \rightarrow B \otimes A$ (the
{\em unit}) and $\epsilon_{A} : A \otimes B \rightarrow I$ (the {\em
co-unit}) such that (here identifying $I \otimes B$ and $B \otimes
I$ with $B$, and suppressing the canonical association morphism $(A
\otimes B) \otimes C \simeq A \otimes (B \otimes C)$)
\[(\id_B \otimes \epsilon_A) \circ (\eta_{A} \otimes \id_{B}) =
\id_{B} .\]
If ${\C}$ is a monoidal category of state spaces (that is, a
sub-category of $\Asp$ equipped with a symmetric, associative
product $A, B \mapsto AB$), then $\eta \in {\C}(I,A \otimes B)$ is
simply a positive, sub-normalized state in $A \otimes B$, while
$\epsilon_{A} \in {\C}(A \otimes B, I)$ is simply an effect on $A
\otimes B$, and we see that this amounts to the definition of a
correction-free teleportation protocol in ${\C}$.

The foregoing discussion suggest one way to make contact between the
structurally loose, but (so to say) ontologically rigid world of
abstract state spaces, and the more highly structured but
ontologically fluid categorical semantics of Abramsky and Coecke:
begin with a particular category of abstract state spaces, and, from
this, construct, either by enlarging it or by paring it down, a
theory having, at a minimum, a sensible duality and monoidal
structure.

It is also worth considering a different, more ``top-down" approach:
to proceed axiomatically, by laying down at the outset a minimum of
constraints on what could count as a category of processes, and
exploring the consequences of further requirements, e.g, that such a
theory support teleportation, or that it not allow cloning, or
bit-commitment. Of course, this is very close to the approach of
Abramsky and Coecke and their collaborators; but we want to suggest
that it may be fruitful to add an extra structural ingredient at the
outset---namely, convexity. This is naturally captured by the
notion of a category of processes {\em enriched over ordered linear
spaces}. In the operational framework, the set of states, the set of
measurement outcomes, the set of dynamics on a system, or more
general operations turning one type of system into another, are all
compact convex sets determined by imposing natural normalization
conditions on convex cones of ``un-normalized'' states, outcomes, or
dynamics, and the convexity is motivated by saying that we should be
able to prepare two states, or perform two measurements, or
implement two dynamics, conditional on the outcome of some random
event with some definite ascribed probabilities $p$ and $1-p$, such
as the flip of a coin with known bias.

In other words, states, outcomes, dynamics, etc. should all belong
to convex cones.  The convexity requirement operationally motivated
above will be implemented in this framework by requiring all
hom-sets in a category describing an operational theory to be
pointed, generating, closed convex cones; more formally, by
requiring a category describing an operational theory to be enriched
over a certain category of ordered linear spaces. This suggests the
following

\begin{definition}
A {\em convex operational category of processes} is a category $\cc$
enriched over the category of finite-dimensional ordered real vector
spaces (with closed, spanning positive cone), with a {\em unit object} 
$I$, such that each object $A$ is
equipped with a distinguished morphism $u_A \in \cc(A, I)$.
\end{definition}

In such a category, states and measurement-outcomes can be
regarded, not as primitives, but as special kinds of processes:
states of a system $A$ are represented  by morphisms from a
distinguished object, the ``unit'' (not to be confused with the
``order unit'' associated with a system), measurement outcomes, by
morphisms from $A$ {\em to } the unit object, and dynamics on system
$A$ by morphisms from $A$ to itself; dynamics changing a system of
type $A$ to one of type $B$ may be represented by morphisms from $A$
to $B$.

In future work, the foregoing definition and its consequences will
be elucidated in more detail. We intend it, and related categorial
formulations of convex operational theories that we and
collaborators are embarked on, to enable the comparison of the
categorial formulation of information-processing centred around
dagger compact closed categories, with the convex operational
formalism.  In the convex formalism, we have been concerned with
obtaining necessary and/or sufficient conditions for the possibility
of particular kinds of information processing, such as the ones we
have reviewed in this paper.  In some cases, it appears these
conditions may be weaker than those employed in existing categorial
constructions.  We hope that the project of categorifying the convex
approach (and convexifying the categorical approach!) may shed more
light on {\em categorically formulated} necessary and sufficient
conditions for various information-processing protocols (about which
much, especially sufficient conditions, is known already), in part
by enabling us to abstract from some of the more concrete content of
the convex formalism while retaining some of its structural
looseness.

\bibliographystyle{plain}

\end{document}